\documentstyle[12pt,epsf,amstex]{article}

\begin{document}
\setlength{\unitlength}{1mm}
\textwidth 15.0 true cm 

\headheight 0 cm
\headsep 0 cm
\topmargin 0.4 true in
\oddsidemargin 0.25 true in
\input epsf

\newcommand{\beq}{\begin{equation}}
\newcommand{\eeq}{\end{equation}}
\newcommand{\be}{\begin{eqnarray}}
\newcommand{\ee}{\end{eqnarray}}
\renewcommand{\vec}[1]{{\bf #1}}
\newcommand{\vecg}[1]{\mbox{\boldmath $#1$}}

\renewcommand{\theequation}{\thesection.\arabic{equation}}

\newcommand{\grpicture}[1]
{
    \begin{center}
        \epsfxsize=200pt
        \epsfysize=0pt
        \vspace{-5mm}
        \parbox{\epsfxsize}{\epsffile{#1.eps}}
        \vspace{5mm}
    \end{center}
}

\begin{flushright}

SUBATECH--01--11\\

\end{flushright}

\vspace{0.5cm}

\begin{center}

{\Large\bf  Quasiclassical expansion for ${\rm Tr} \{ (-1)^F e^{-\beta H} \} $.}

\bigskip

   {\Large  A.V. Smilga} \\

\vspace{0.8cm}

{\it SUBATECH, Universit\'e de
Nantes,  4 rue Alfred Kastler, BP 20722, Nantes  44307, France. }\\

\end{center}

\bigskip

\begin{abstract}
We start with some methodic remarks referring to purely bosonic
quantum systems  and then explain how 
corrections to the leading--order quasiclassical 
result for 
 the fermion--graded partition function 
${\rm Tr} \{ (-1)^F e^{-\beta \hat H} \} $ can be calculated at small 
$\beta$. We perform such calculation
  for certain  supersymmetric 
quantum mechanical systems where such corrections are expected to appear.
We consider in particular  supersymmetric Yang-Mills 
theory reduced to
(0+1) dimensions and were surprised to find that  the correction 
$\propto \beta^2 $ vanishes in this case.
 We discuss also a
nonstandard ${\cal N} =2$ supersymmetric $\sigma$--model
defined on  $S^3$ and other 
3--dimensional conformally flat manifolds and 
show that the
quasiclassical expansion breaks down for this system. 

\end{abstract}

\section{Introduction}
For many supersymmetric quantum systems, the fermion--graded partition function
\footnote{This is how the object (\ref{ZSUSY}) will be called in this paper
 (no established term for it exists in the literature). We admit that this 
name is awkward, but it is not misleading like ``supersymmetric partition function"
 or ``index" would be. {\it (i)}
$Z^{F{\rm-grade}} (\beta)$ {\it is} not the partition function 
${\rm Tr} \{ e^{-\beta \hat H} \}$ of a supersymmetric system.  
Besides, though supersymmetric 
theories provide the main motivation and interest in studying the 
quantity  
(\ref{ZSUSY}), 
the latter can be defined for {\it any} system involving fermion degrees of 
freedom. 
{\it (ii)}  If $Z^{F{\rm-grade}} (\beta)$  depends on $\beta$, it does not 
coincide 
with the index (\ref{IWit}). }
\be
\label{ZSUSY}
Z^{F{\rm-grade}} (\beta) \ =\ {\rm Tr} \{ (-1)^F e^{-\beta \hat 
H} \} 
 \ee
does not depend on $\beta$ and and defines the Witten index of the system
 \be
\label{IWit}
I_W \ =\ n_B^{(0)} - n_F^{(0)} \ =\ \lim_{\beta \to \infty} 
Z^{F{\rm-grade}} (\beta) \ .
 \ee
Usually, $Z^{F{\rm-grade}} (\beta)$ can be evaluated at small $\beta$ with
quasiclassical methods. The leading order result is \cite{Cecotti} 
  \be
 \label{intCec}
Z^{F{\rm-grade}} (\beta) \ =\ \int \prod_i \frac {dp_i dq_i}{2\pi}
\prod_a d\bar\psi^a d\psi^a \exp \left\{ - 
\beta H^{\rm cl} (p_i, q_i; \bar \psi^a, \psi^a) \right\} \ ,
  \ee
where $(p_i, q_i)$ and $(\bar\psi^a, \psi^a)$ are bosonic and fermionic
phase space variables, and $H^{\rm cl}$ is the classical Hamiltonian function
corresponding to the quantum Hamiltonian $\hat H$.

This philosophy does not always work, however. First of all, one can claim
that $Z^{F{\rm-grade}} (\beta)$ is $\beta$--independent only for the systems
with discrete spectrum. Then the contributions of the 
degenerate boson and fermion states cancel in the trace and 
only the 
zero--energy states contribute. If the spectrum is continuous, 
it is not immediately clear what the trace or supertrace {\it is}. Some 
regularization is required or an additional definition for the quantity
$Z^{F{\rm-grade}} (\beta)$ should be given. After that
 $Z^{F{\rm-grade}} (\beta)$ can well display 
 a nontrivial dependence on $\beta$. 

As an
important example of supersymmetric quantum mechanics (SQM)
 with continuous spectrum, consider the 
super-Yang-Mills (SYM) quantum mechanics
 obtained by dimensional reduction
of  SYM theories. Consider first the system obtained from   
${\cal N} =1,
d=4$ SYM theory based on the gauge group $G$. The Hamiltonian 
has the form 
  \be
\label{hamSYM}
\hat H \ =\ \frac 12 \hat P_i^a  \hat P_i^a + \frac {g^2}4 f^{abe} f^{cde}
A_i^a A_j^b A_i^c A_j^d + ig f^{abc} \hat{\bar\lambda}^{a\alpha} 
(\sigma_i)_\alpha^{\ \beta}\lambda^b_\beta A^c_i \ , \nonumber \\
i = 1,2,3,\ \ \ \   \alpha = 1,2\ \ \ \ \ \ \ \ a = 1, \ldots , 
{\rm dim}(G) \ .
 \ee
$A_i^a$ are the gauge potentials, $\hat P_i^a \equiv \hat E_i^a = 
-i\partial/\partial A_i^a$ 
are their canonical momenta operators, and 
$\lambda^a_\alpha$ and 
$\hat{\bar\lambda}^{a\alpha} = \partial/\partial \lambda^a_\alpha $
are the fermionic gluino variables and their momenta. 
The Hilbert
space includes only the physical
states annihilated by the Gauss law constraints
 \be
   \label{conSYM}
  \hat G^a \Psi \ =\ f^{abc} \left( \hat P_i^b A_i^c + i 
\hat {\bar \lambda}^{b\alpha} 
\lambda_\alpha^c \right) \Psi \ =\ 0\ .
 \ee 
The system has two conserved
complex supercharges
 \be
\label{QSYM}
\hat Q_\alpha \ =\  \frac 1{\sqrt{2} }  (\sigma_i)_\alpha^{\ \beta} 
\lambda_\beta^a \left[\hat E^a_i +
\frac {ig}2 \epsilon_{ijk} f^{abc} A^b_j A^c_k \right]
 \ee
(they formed a Weyl spinor before reduction) and, being restricted
on the Hilbert space (\ref{conSYM})  enjoys the ${\cal N} =2$
SQM algebra $\{ \hat{\bar Q}^\alpha, \hat Q_\beta \}_+ = 
\delta^\alpha_\beta \hat H $.
The classical potential in Eq.(\ref{hamSYM}) vanishes in the ``vacuum valleys''
with $f^{abc}A_i^b A_j^c = 0$. Due to supersymmetry, degeneracy
along the valleys survives also after quantum corrections are taken into 
account. As a result, the system tends to escape
along the valleys, the wave function of the low--energy states 
is delocalized, and the 
spectrum is continuous (this implies, incidentally, the continuity of the
mass spectrum of supermembranes \cite{contspec}). The calculation of 
$Z^{F{\rm-grade}} (\beta)$ for the system (\ref{hamSYM}) in the quasiclassical 
aproximation gives a fractional number.
 For example, for the $SU(2)$ gauge group \cite{jaIW,Sethi,Kac}
 \be
\label{Zeq14}
\lim_{\beta \to 0} Z^{F{\rm-grade}} (\beta) \ =\ \frac 14 \ .
 \ee
On the other hand, this system does not have  normalized  vacuum 
state with zero energy and 
 \be
\label{IWnol}
I_W \ =\ \lim_{\beta \to \infty} Z^{F{\rm-grade}} (\beta) 
\ =\ 0 \ .
 \ee
A similar mismatch between the values of $ Z^{F{\rm-grade}} (\beta)$ in the two
limits shows up in more complicated SYM QM systems. $ Z^{F{\rm-grade}} (0)$
is always fractional \cite{Nekr,Stau}, while the Witten index vanishes for the 
${\cal N} =2$ and ${\cal N} =4$ systems (the latter are obtained by dimensional
reduction from 6--dimensional SYM theories) and acquires a nonzero integer 
value for the systems obtained by the reduction of 10--dimensional SYM 
theories and involving 8 complex supercharges \cite{Halpern,Hoppe,Kac}.

In Sect. 5 we will deal with this system and calculate the
1--loop correction
to the fractional leading order result for 
$ Z^{F{\rm-grade}} (\beta)$ at small
$\beta$. Remarkably, the correction  of order $\beta^2$
 {\it vanishes} in all cases.  We expect, however, nontrivial
corrections  to appear in higher loops. 

For the systems with discrete spectrum,   $ Z^{F{\rm-grade}} (\beta)$  
is $\beta$--independent, but, contrary to naive expectations, it 
does {\it not} always mean that it can be evaluated with the quasiclassical
formula (\ref{intCec}). In Sect. 6 we discuss a rather nontrivial example
of a nonstandard ${\cal N} =2$ supersymmetric $\sigma$ 
model living on $S^3$, for 
which the phase space integral (\ref{intCec}) does not give a correct 
result for the index.

The model belongs to a class of ${\cal N} =2$ supersymmetric $\sigma$ 
models living on 3--dimensional conformally flat manifolds \cite{Heff,Ivanov}.
The supercharges and the Hamiltonian of the model have the form
\be
\label{QHN2}
\hat Q_\alpha \ =\ - i \sqrt{\frac 12} \left [ (\sigma_k)_\alpha^{\ \beta} 
\psi_\beta f(\vec{x}) \hat p_k + 
i \partial_k f(\vec{x}) 
\hat {\bar \psi} \sigma_k \psi  \psi_\alpha  
\right] \ , \nonumber \\
  \hat {\bar Q}^\alpha \ =\  i \sqrt{\frac 12} \left [\hat {\bar \psi}^\beta 
(\sigma_k)_\beta^{\ \alpha}
f(\vec{x}) \hat p_k -
i \partial_k f(\vec{x})  \hat {\bar\psi} \sigma_k \psi \hat 
{\bar \psi}^\alpha  
\right] \ , \nonumber \\
\hat H \ = \  \frac 12 f(\vec{x})  \hat p_k^2 f(\vec{x})
 - \epsilon_{jkp} \hat {\bar\psi} \sigma_j \psi  f(\vec{x})
\partial_p f(\vec{x}) \hat p_k
- \frac 12  f(\vec{x}) \partial^2_k f(\vec{x})
(\hat {\bar\psi} \psi )^2 \ ,
 \ee
where the differential operators $\hat p_k = -i\partial/\partial x_k$ and
$\hat {\bar\psi}_\alpha = \partial/\partial \psi_\alpha$ 
act on everything on the right they find. The algebra 
$$ \left\{ \hat  Q_\alpha,  \hat  Q_\beta \right\}_+ = 0\ ,
\left\{ \hat {\bar Q}^\alpha,  \hat  Q_\beta \right\}_+  = 
\delta^\alpha_\beta \hat H 
 $$ 
holds.

If the manifold is compact, the spectrum is discrete. For $S^3$ 
[with $f(\vec{x}) = 1+\vec{x}^2/(4R^2)$ ], there are two 
bosonic states with
zero energy, which gives $I_W = 2$. On the other hand, the 
integral (\ref{intCec})
is not equal to 2. Moreover, its value  depends on the way 
the ordering ambiguities are resolved and a classical counterpart of the
quantum Hamiltonian in Eq.(\ref{QHN2}) is defined.
Also, there are nonvanishing corrections, 
which are of the same order
as the leading order result and as the two--loop and higher--loop 
corrections.
The whole
quasiclassical expansion for $ Z^{F{\rm-grade}} (\beta)$ breaks down.

To prepare ourselves for the discussion of these comparatively complicated
problems, we start in the next section with recalling the technique of 
calculating the loop corrections  to the ordinary partition function
$Z = {\rm Tr} \exp\{-\beta \hat H\}$ for purely bosonic systems. 
Gauge QM systems, where the calculation 
involves certain 
subtleties, are considered in Sect. 3. 
In Sect. 4 we generalize the analysis to the systems with fermion
dynamic variables, being especially interested  in supersymmetric systems.
We consider the simplest SQM system due to Witten \cite{WitSQM}, where 
the corrections to  $ Z^{F{\rm-grade}} (\beta)$ vanish even in the
cases when the spectrum is continuous and one could expect
{\it a priori} the corrections to appear. 
Sect. 5 is devoted to
SYM QM and Sect. 6 --- to the  ${\cal N} =2$ supersymmetric 
$\sigma$--model
on $S^3$. The last section is reserved, as usual, to recapitulating
the results and to the discussion.

\section{Quasiclassical expansion of the partition function.}
\setcounter{equation}0

Consider a purely bosonic QM system. To  leading order, the partition
function is given by the integral
 \be
\label{Zboson}
Z \ =\ {\rm Tr} \{ e^{-\beta \hat H}\} \ =\ 
 \int \prod_i \frac {dp_i dq_i}{2\pi}
 \exp \left\{ - 
\beta H^{\rm cl} (p_i, q_i) \right\} \ .
 \ee
Strictly speaking, the function $H^{\rm cl}$ is not uniquely defined due to
the ordering ambiguities, but the ambiguity in the prescription 
$ \hat H \to  H^{\rm cl} (p_i, q_i)$ does not affect the results in the 
leading order in $\beta$.
\footnote{This is true for $Z$, but not always true for 
$Z^{F{\rm-grade}}$. See Sect. 6.}
We will stick to the most convenient choice and assume that 
$ H^{\rm cl} (p_i, q_i)$ coincides with the {\it Weyl symbol} \cite{Weyl}
of the quantum Hamiltonian $\hat H$. For sure, the final results (when all 
corrections are taken into account) do not depend
on convention.

The result (\ref{Zboson}) represents the leading term in the high-temperature
(small $\beta$) expansion for $Z$. The next-to-leading term is suppressed
compared to Eq.(\ref{Zboson}) 
as $\sim \beta^2 E_{\rm char}^2$, where $E_{\rm char}$ is the
characteristic energy scale. For the  Hamiltonian of the 
simplest type,
 \be
\label{Hsimple}
H(p_i, q_i) \ =\ \frac {p_i^2}2 + V(q_i)\ ,
 \ee
the partition function is given by the well-known expression \cite{Feyn}
   \be
\label{corrZ}
Z \ =\  
 \int \prod_i \frac {dp_i dq_i}{2\pi}
 e^{ - \beta H (p_i, q_i) } 
\left[ 1 - \frac {\beta^2}{24} \frac {\partial^2 V}{\partial q_i^2}
+ O(\beta^4) \right]
 \ee
($H(p_i,q_i) \equiv H^{\rm cl}(p_i,q_i)$).  
In a generic case, the result is \cite{Robert} 
  \be
\label{corrZHpq}
Z \ =\ \int \prod_i \frac {dp_i dq_i}{2\pi}
 e^{ - \beta H (p_i, q_i)} 
\left[ 1 - \frac {\beta^2}{24} \left(\frac {\partial^2 H}{\partial p_i 
\partial p_j}
\frac {\partial^2 H}{\partial q_i \partial q_j} \right. \right. 
\nonumber \\
\left. \left. -\  
 \frac {\partial^2 H}{\partial p_i \partial q_j} 
 \frac {\partial^2 H}{\partial p_j \partial q_i} \right) 
+ O(\beta^4) \right]\ .
 \ee

The result (\ref{corrZHpq}) can be derived in two ways. First, one can note
that 
  \be
\label{ZexpWeyl}
Z \ =\  
 \int \prod_i \frac {dp_i dq_i}{2\pi}
\left[ e^{ - \beta \hat H} \right]_W \ ,
  \ee
where $[\hat O]_W$ is the Weyl symbol of the operator $\hat O$. To
 leading
order, the Weyl symbol of the exponential is given by the exponential of
the Weyl symbol [this gives Eq.(\ref{Zboson})], but there are corrections.
In the modern language, they stem from the fact that the ``star product''
is not just a simple product.
 In the Appendix, we present, following Ref.\cite{Robert},
an accurate calculation of such corrections, which leads to 
Eq.(\ref{corrZHpq}).

Alternatively, one can use 
the functional integral representation for the partition function. 
To make the problem simpler, consider first the case with
 only one pair $(p,q)$
of phase space variables. The path integral for the partition function is
 \be
\label{Zpathbos}
Z \ =\ \int {\cal D} p  {\cal D} q \exp \left\{ \int_0^\beta d\tau
[ip\dot q - H(p,q)] \right\} \ ,
  \ee
where $\tau$ is the Euclidean time. Periodic boundary conditions
  \be
\label{bcboson}
q(\beta) = q(0),\ \ \ \ p(\beta) = p(0)
  \ee
are imposed. For very small $\beta$, we can ignore the $\tau$--dependence of
our variables and set $q(\tau) \approx \bar q$ and $p(\tau) 
\approx \bar p$.
The measure ${\cal D} p  {\cal D} q$ happens to go into 
$(d\bar p d\bar q)/(2\pi)$ and we are reproducing the result (\ref{Zboson}).
To find the corrections, we write
  \be
\label{zamena}  
q(\tau) \ =\ \bar q + x(\tau),\ \ \ \ \ \ p(\tau) 
\ =\  \bar p + s(\tau) 
  \ee
with $\int d\tau x(\tau) = \int d\tau s(\tau) = 0$ and assume $x(\tau)$
and $s(\tau)$ to be small. Expanding over $x(\tau)$
and $s(\tau)$ up to the second order (the linear terms vanish), we obtain 
 \be
\label{Zsx}
Z  \approx \int  \frac {d\bar p d\bar q}{2\pi} 
 e^{ - \beta H (\bar p, \bar q)}
\int  {\cal D}' s  {\cal D}' x  \nonumber \\
\exp \left\{ \int_0^\beta d\tau  
\left[ i s(\tau) \dot x(\tau)  
  - \frac A2  s^2(\tau) -
 \frac C2  x^2(\tau) -  B s(\tau) x(\tau)
\right] \right\} \ ,
 \ee
where
  \be
 \label{ABC}
A =   \frac {\partial^2 H}{\partial \bar p^2},\ \ \ 
B =   \frac  {\partial^2 H}{\partial \bar p \partial \bar q}, \ \ \ 
C =   \frac {\partial^2 H}{\partial \bar q^2} \ .
  \ee
The prime in the measure $ {\cal D}' s  {\cal D}' x$ means 
that the zero Fourier harmonics of $s(\tau)$ and $x(\tau)$ are
constrained to be zero. 
Let us make now the canonical transformation
  \be
  \label{canontr}
S \ =\ A^{1/2} \left(s + \frac{Bx}A\right),\ \ \ \ \ 
X \ =\ A^{-1/2} x \ \ \ .
  \ee
It leaves invariant the measure and the Poisson bracket,
 \be
\label{PB}
\{ S(\tau), X(\tau')\}_{P.B.} \ =\ \{ s(\tau), x(\tau')\}_{P.B.} 
\ =\ \delta(\tau - \tau') - \frac 1\beta \ .
  \ee
We obtain 
 \be
\label{ZSX}
Z \ =\ \int  \frac {d\bar p d\bar q}{2\pi} 
 e^{ - \beta H (\bar p, \bar q)}
\int  {\cal D}' S  {\cal D}' X \nonumber \\  
\exp \left\{ \int_0^\beta d\tau  
\left[ i S(\tau) \dot X(\tau)  
  - \frac 12 S^2(\tau) -
 \frac 12 
 \omega^2_{\bar p \bar q}  X^2(\tau)
\right] \right\} \ ,
 \ee
where
  \be
\label{omega}
 \omega^2_{\bar p \bar q} \ =\ 
\frac {\partial^2 H}{\partial \bar p^2}   
 \frac {\partial^2 H}{\partial \bar q^2} - 
\left(\frac {\partial^2 H}{\partial \bar p \partial \bar q} 
\right)^2 \ . 
  \ee
If the integration over the zero harmonics of 
$S(\tau)$ and $X(\tau)$ were also included,
the inner integral would determine the partition function of
the harmonic oscillator. As it is not included, the integral
represents the ratio of the full partition function of the
oscillator
  \be
  \label{Zosc}
Z^{\rm osc} \ =\ \int  {\cal D} S  {\cal D} X  
\exp \left\{ \int_0^\beta d\tau  
\left[ i S(\tau) \dot X(\tau)  
  - \frac 12 S^2(\tau) -
 \frac 12  \omega^2  X^2(\tau)
\right] \right\} \nonumber \\
 =\ \frac 1 {2 \sinh \left( \frac {\beta\omega}2 \right)}
  \ee
and the integral 
  \be
  \label{Zoscquas}
\int  \frac{d \bar S  d \bar X}{2\pi}  
\exp \left\{ - \frac \beta 2
[  \bar S^2 +
   \omega^2  \bar X^2 ] \right \} \ =\ \frac 1 {\beta\omega} \ .
  \ee
[Eq.(\ref{Zoscquas}) is none other than the partition function of the
oscillator in the quasiclassical limit]. We finally obtain 
 \be
\label{Z1para}
Z \ \approx\ \int  \frac {d\bar p d\bar q}{2\pi} 
 e^{ - \beta H (\bar p, \bar q)}
\frac  {\beta\omega_{\bar p\bar q}} {2 \sinh \left( 
\frac {\beta\omega_{\bar p \bar q}}2 \right)} \nonumber \\
\approx \int  \frac {d\bar p d\bar q}{2\pi} 
 e^{ - \beta H (\bar p, \bar q)}
\left[1 - \frac {\beta^2}{24} \omega^2_{\bar p \bar q} + \cdots
\right]
 \ee
in accordance with Eq.(\ref{ZexpWeyl}). The higher--order terms of
the expansion of $ \sinh^{-1} (\beta\omega/2 )$ give the corrections
$\propto \beta^4$ etc, but, to take the latter
 into account correctly, one
has also to keep higher terms in the expansion of the Hamiltonian
over $s(\tau)$ and $x(\tau)$.

In the general multidimensional case, 
the corrections may be found by solving the quantum mechanical
problem with the Hamiltonian 
  \be
\label{HPauli}
\tilde H \ =\  \frac 12 A_{jk}  s_j  s_k  +  B_{jk} s_j x_k +
 \frac 12 C_{jk} x_j x_k\ ,
 \ee
$$ A_{jk} =  
 \frac {\partial^2 H}{\partial \bar p_j \partial \bar p_k },\ \ \ 
B_{jk} =  \frac  {\partial^2 H}{\partial \bar p_j 
\partial \bar q_k}, \ \ \ 
C_{jk} =   \frac {\partial^2 H}{\partial \bar q_j  
\partial \bar q_k } \ .
$$
With a proper choice of variables, 
the Hamiltonian (\ref{HPauli}) describes the (multidimensional)
motion of a scalar charged particle in a generic oscillator
potential and in a constant magnetic field. The problem can
be solved exactly. 
  It is reduced to a set of
oscillators whose frequencies can be found algebraically. The simplest way 
to do it is 
to  calculate the Gaussian path integral for
$Z$. On the first step, we integrate over the momenta $s_j(\tau)$
(with the zero Fourier modes subtracted)
and express  the multidimensional analog of the inner
integral in  Eq.(\ref{Zsx})  as
   \be
I \ \propto\ \prod_j \int {\cal D}'x_j \exp 
\left\{- \int_0^\beta {\cal L}_E \right \} \ ,
   \ee
where
 \be
 \label{LE} 
{\cal L}_E \ =\ \frac 12  (A^{-1})_{jk} \dot x_j \dot x_k
- i (A^{-1} B)_{jk} \dot x_i x_k + \frac 12 (C - B^T A^{-1} B)_{jk} x_j x_k \ .
  \ee
 Let us expand 
 \be
\label{sxexpan}
x_j(\tau) \ =\ \sum_{n=1}^\infty x_j^{(n)} e^{2\pi in\tau/\beta}
\ +\ {\rm complex \ conjugate} \ 
 \ee
and do the integral over 
$$ \prod_j {\cal D}'x_j \ \propto\ \prod_j \prod_{n=1}^\infty 
d x_j^{(n)} d \bar x_j^{(n)} \ .$$
We obtain
\be 
\label{otnprod}
I \ =\ \prod_{n=1}^\infty 
\frac{\det\| A^{-1} \omega_n^2 \|}
{\det\| A^{-1} \omega_n^2  + (A^{-1}B - B^TA^{-1})\omega_n + C - 
B^T A^{-1} B \|} = \nonumber \\
\prod_{n=1}^\infty  \frac{\omega^{2N}}
{\det\|  \omega_n^2  + (B - AB^TA^{-1})\omega_n + AC - 
A B^T A^{-1} B \|} \ ,
 \ee
where 
 \be
\label{omegan}
\omega_n = \frac {2\pi n}\beta 
 \ee
and $N$ is the number of degrees of freedom. 
The normalization factor in Eq.(\ref{otnprod}) is chosen such
that $I$ represents the ratio of the full partition function of the
system (\ref{HPauli}) and this partition function in the 
quasiclassical limit. For the free Hamiltonian $H = (1/2)A_{jk}p_j 
p_k$, $I = 1$. 

Now, we write
\be
\label{detOm}
\det\|  \omega_n^2  + (B - A B^TA^{-1})\omega_n + AC - 
AB^T A^{-1} B \| = \prod_{j=1}^N (\omega_n^2 + \Omega_j^2) \ ,
 \ee
where $-\Omega_j^2$ are the roots of the corresponding polynomial.
\footnote{To prove that the left side of Eq.(\ref{detOm}) is, indeed, a polynomial
in $\omega_n^2$, let us choose a basis where $A= 1\!\!\!1$. We are left with
the expression $\det \|\omega_n^2 + \omega_n F + S\|$, where
$S$ and $F$ are symmetric and antisymmetric real matrices. It is not difficult 
to see now that the   odd powers of $\omega_n$
in the expansion of the determinant cancel out.}

The ratio (\ref{otnprod}) acquires the form 
 \be 
\label{IOm}
I \ =\ \prod_{j=1}^N \left( 
\prod_{n=1}^\infty \frac {\omega_n^2}{\omega_n^2 + \Omega_j^2}
\right) =  \ \prod_{j=1}^N 
\frac {\beta \Omega_j}{2 \sin \frac {\beta \Omega_j}2 }\ .
 \ee
The full partition function
   \be 
\label{ZOm}
Z \ =\ \prod_{j=1}^N 
\frac {1}{2 \sin \frac {\beta \Omega_j}2 }
 \ee
represents a product of the partition functions of the harmonic oscillators
with frequencies $\Omega_j$. Generically,  $\Omega_j$ appearing on the
right side of Eq.(\ref{detOm}) are complex.
For example, it is so for a not positive definite and hence
 non-Hermitian Hamiltonian (\ref{HPauli}) with 
$A = -C = 1\!\!\!1$, $B=0$.
      If the Hamiltonian is Hermitian, all   $\Omega_j$ are real. The spectrum
has the form
  \be
 \label{spectr}
E_{\{n_j\}} \ =\ \sum_{j=1}^N \left( \frac 12 + n_j \Omega_j \right)\ .
 \ee

As a simplest nontrivial example, consider the 2--dimensional Hamiltonian
 \be
\label{HH}
H \ = \frac 12 \left( p_x - \frac {Hy}2 \right)^2 + 
\frac 12 \left( p_y + \frac {Hx}2 \right)^2 + \frac 12 \left(
\omega_1^2 x^2 + \omega_2^2 y^2 \right) \ .
  \ee
 Eq.(\ref{detOm}) acquires the form
 \be
\label{detH2}
\det \left\| \begin{array}{cc}
\omega_n^2 + \omega_1^2 & \omega_n H \\
-  \omega_n H & \omega_n^2 + \omega_2^2  \end{array} \right\| \ 
= (\omega_n^2 + \Omega_1^2) (\omega_n^2 + \Omega_2^2)\ ,
  \ee
which gives the frequencies
 \be
\label{Om12}
 \Omega_{1,2} \ =\ 
\frac 12 \left[ \sqrt{H^2 + (\omega_1 + \omega_2)^2} 
\pm  \sqrt{H^2 + (\omega_1 - \omega_2)^2} \right]\ .
  \ee
The result (\ref{Om12}) was obtained earlier by a different method
\cite{velo}.

 In this paper, our primary concern are  the corrections 
$\propto \beta^2$ in the partition function. To find them,  we
do not need to determine all eigenfrequencies $\Omega_j$.
It suffices to expand the right side of Eq.(\ref{otnprod})
in $\beta$ 
using the identity
$$\det \|1 + \alpha \| \ =\ 1 + {\rm Tr}\ \alpha + 
\frac 12 [({\rm Tr}\ \alpha)^2 - {\rm Tr} \ \alpha^2 ] 
+ o(\alpha^2)\ .$$
We  obtain
 \be
\label{otnotv}
I \ \approx \ \prod_{n=1}^\infty 
\frac{ \omega_n^2 }
{\omega_n^2 + {\rm Tr} (AC - B^2)} \ \approx \ 
1 - \frac {\beta^2}{24} {\rm Tr} (AC - B^2)\ ,
 \ee
where the relation 
$\sum_{n=1}^\infty (1/n^2) = \pi^2/6$ was used. The result
(\ref{otnotv}) coincides with the square bracket
in Eq.(\ref{corrZHpq}). 

\section{Gauge quantum mechanics.}
\setcounter{equation}0

In order to prepare ourselves for the discussion of the
SYM quantum mechanics, we are going to be as instructive 
as possible and consider two simple quantum mechanical models
involving gauge symmetry.

\subsection{$SO(2)$ gauge oscillator.}

The simplest gauge QM system is 
the constrained two--dimensional oscillator
(see e.g. Ref.\cite{Annals,kniga}). It is the system 
described by the Hamiltonian
 \be
\label{HO2}
 \hat H = \ \frac 12 \hat p_i^2 + \frac 12 \omega^2 x_i^2 
 \ee
with the constraint 
 \be
\label{ConO2} 
\hat p_\phi \Psi^{\rm phys} \ =\ \epsilon_{jk}x_j \hat p_k
\Psi^{\rm phys} \ =\ 0 \ .
 \ee
The spectrum of the system involves all rotationally invariant 
states $\Psi^{\rm phys}_n$,
$\epsilon_n = \omega(1+2n)$, and the partition function can be found
straightforwardly
 \be
 \label{ZO2}
Z^{O(2)} (\beta) \ =\ \sum_n e^{-\beta \epsilon_n} \ =\ 
\frac1{2\sinh(\beta\omega)} \ .
  \ee
The same result is obtained by using  functional methods. We 
start from the representation
  \be
\label{ZK}
Z^{O(2)}(\beta) \ =\ \int_0^{2\pi} \frac {d\phi}{2\pi}
\int d\vec{x}\ {\cal K}(\vec{x}^\phi, \vec{x}; -i\beta) \ ,
  \ee
where $ {\cal K}(\vec{x}^\phi, \vec{x}; -i\beta)$ is the Euclidean
evolution operator of the unconstrained system (\ref{HO2}) and
$x_i^\phi = O_{ij}(\phi) x_j$, $O_{ij}$ being an $SO(2)$ matrix.
The integral over $d\phi$ kills all rotationally noninvariant states in
the spectral decomposition of  $ {\cal K}$. Eq.(\ref{ZK}) can be expressed
into a path integral,
  \be
\label{ZO2path}
Z^{O(2)}(\beta) \ =\ \int_0^{2\pi} \frac {d\phi}{2\pi}
\int \prod_\tau  d\vec{x}(\tau) \exp\left\{
- \int_0^\beta d\tau {\cal L}_E[\vec{x}(\tau)] \right \}\ ,
 \ee
where    
\be
\label{LEbc}
 {\cal L}_E\ &=&\ \frac 12 \dot x_i^2 + \frac 12 
\omega^2 x_i^2 \ , \nonumber \\
\vec{x}(\beta) \ &=&\ \vec{x}^\phi(0)\ .
  \ee
The boundary conditions can be rendered periodic by changing the variables
\be
 \vec{x}(\tau) \ =\ \vec{y}^{\phi\tau/\beta}(\tau) \ .
 \ee
We arrive at 
    \be
\label{ZO2pathy}
Z^{O(2)}(\beta) \ =\ \int_0^{2\pi} \frac {d\phi}{2\pi}
\int \prod_\tau  d\vec{y}(\tau) \exp\left\{
- \int_0^\beta d\tau \tilde{\cal L}_E[\vec{y}(\tau)] \right \}\ ,
 \ee
where 
  \be
\label{LEO2t}
 \tilde {\cal L}_E\ =\ \frac 12 \left(\dot y_i + 
\frac \phi\beta \epsilon_{ij}y_j \right)^2 +
\frac 12 
\omega^2 y_i^2
  \ee
and $\vec{y}(\beta) = \vec{y}(0)$. If one wishes, one can upgrade the integral
over $d\phi$ to the path integral over the periodic functions $\phi(\tau)$.
Then the Lagrangian (\ref{LEO2t}) with time--dependent  $\phi(\tau)$ 
is invariant under the gauge transformations
  \be
 y_i \ =\ O_{ij}(\chi) y'_j, \ \ \ \ \ \ \phi \ =\ \phi' - 
\beta \dot \chi \ .
 \ee
The form (\ref{ZO2pathy}) is more convenient, however. Substituting there the
Fourier expansion for $\vec{y}(\tau)$ and doing the Gaussian integrals,
we obtain
  \be
\label{ZO2prod}
Z^{O(2)}(\beta) \ =\ \int_{-\pi}^{\pi} \frac {d\phi}{2\pi}
\frac 1{(\beta\omega)^2 + \phi^2} \prod_n' 
\frac {\omega_n^2} {(\omega_n + \phi/\beta )^2 + \omega^2}
 \ ,
 \ee
where the product $\prod'_n$ is done over positive and negative nonzero
integer $n$. The integrand is periodic in $\phi$ and we used this
fact, while changing the integration limits. The constant factor is fixed by the
requirement that the inner integral in  Eq.(\ref{ZO2pathy}) reproduces in
the limit $\phi = 0$ the partition function of the unconstrained 
2--dimensional oscillator. Performing the product, we obtain
   \be
\label{chsh}
Z^{O(2)}(\beta) \ =\ \int_{-\pi}^{\pi} \frac {d\phi}{2\pi}
\frac 1{2[\cosh(\beta\omega) - \cos \phi]} \ =\ 
\frac 1{2 \sinh(\beta\omega)} \ .
 \ee
Let us now study the quasiclassical expansion of $Z$. To leading order,
the product over nonzero modes in  Eq.(\ref{ZO2prod}) can be ignored,
and we have
 \be
\label{ZO2q}
[Z^{O(2)}(\beta)]^{\rm quasicl}  \approx 
\int_{-\pi}^{\pi} \frac {d\phi}{2\pi}
\frac 1{(\beta\omega)^2 + \phi^2} \approx 
\int_{-\infty}^{\infty} \frac {d\phi}{2\pi}
\frac 1{(\beta\omega)^2 + \phi^2} = \frac 1{2\beta\omega} \ .
 \ee
The same result is obtained in the Hamiltonian approach
 \be
\label{ZO2qint}
[Z^{O(2)}(\beta)]^{\rm quasicl}  =
\int \prod_{i=1}^2 
\frac {dp_i dx_i}{2\pi} \delta(p_\phi) \exp\{-\beta H\} \nonumber \\
 = \int_{-\infty}^{\infty} \frac {d\phi}{2\pi}
\int \prod_{i=1}^2 
\frac {dp_i dx_i}{2\pi} \exp\{-\beta \tilde H\}
\ ,
 \ee
where 
 \be 
\label{HO2t}
\tilde H \ =\ H + \frac {i\phi}\beta p_\phi \ .
  \ee
To find the higher order terms in the quasiclassic expansion, let us expand
the integrand in Eq.(\ref{ZO2prod}) [or in Eq.(\ref{chsh}), which is
easier] over $\phi$ and $\beta\omega$. Taking into account the leading
and the next--to--leading term, we obtain
  \be
 \label{expan}
 Z  \approx 
\int \frac {d\phi}{2\pi}
\frac 1{(\beta\omega)^2 + \phi^2} 
\left[ 1 - \frac {(\beta\omega)^2 - \phi^2}{12} + \ldots \right] \ .
 \ee
 We seem to be in hot water now: the integral over $\phi$ diverges
linearly at large $\phi$ or, if we keep the 
integration limits finite, it
is determined by  the region of large $\phi$, where the expansion is
not valid. Still, the correction can be calculated with the following recipe:
\begin{itemize}
\item Extend the limits of integration over $\phi$  to $\pm \infty$
\item Forget about the divergence and calculate the correction as the
residue of the integrand at the pole at $\phi = i\beta\omega$ .
 \end{itemize}
We obtain, indeed
  \be
 \label{ZO2expan}
 Z^{O(2)}(\beta)  \approx \ \frac 1{2\beta\omega} 
\left[ 1 - \frac {(\beta\omega)^2}6 + \ldots \right] \ ,
 \ee
which coincides with the expansion of Eq.(\ref{ZO2}). Also 
next--to--next--to--leading and all other corrections 
can be obtained in this way.

This recipe seems to be rather wild, but it is not difficult to 
justify it
quite rigourously. This is done with the following chain of 
relations
   \be
Z^{O(2)}(\beta)  = \int_{-\pi}^{\pi} \frac {d\phi}{4\pi}
\frac 1{[\cosh(\beta\omega) - \cos \phi]} \nonumber \\
 =  \lim_{N\to \infty}
\frac 1{1+2N} \int_{-\pi(2N + 1)}^{\pi (2N +1)} \frac {d\phi}{4\pi}
\frac 1{[\cosh(\beta\omega) - \cos \phi]} \nonumber \\
\approx \lim_{N\to \infty}
\frac 1{1+2N} 2\pi i \sum_{k = -N}^N \ {\rm 
res}_{\phi = i\beta\omega + 2k\pi}
\frac 1{4\pi[\cosh(\beta\omega) - \cos \phi]} = \nonumber \\
2\pi i \ {\rm res}_{\phi = i\beta\omega}
\frac 1{4\pi[\cosh(\beta\omega) - \cos \phi]}  \nonumber \\  
 = i\ {\rm res}_{\phi = i\beta\omega}\left\{ 
\frac 1{(\beta\omega)^2 + \phi^2} \left[ 1 - \frac 
{(\beta\omega)^2 - \phi^2}{12} 
+ \ldots \right] \right\}
 \ .
 \ee
 We want to note here that essentially the same procedure of replacing 
divergent integrals by the residue contributions was used in 
Refs.\cite{Nekr,Stau} when calculating the fermion--graded partition
function for SYM QM in the leading quasiclassical approximation. No
justification of this procedure was given there, but we believe that
it can be eventually found along the same lines as in the trivial
example discussed above. 

The final remark is that the corrections $\propto (\beta\omega)^2$
and $\propto \phi^2$ in  the expansion (\ref{expan}) can, in the
full analogy with Eq.(\ref{corrZHpq}),  be cast in the form
    \be
\label{corrHt}
 - \frac {\beta^2}{24} \left(\frac {\partial^2 \tilde H}{\partial p_i 
\partial p_j}
\frac {\partial^2 \tilde H}{\partial q_i \partial q_j} - 
 \frac {\partial^2 \tilde H}{\partial p_i \partial q_j} 
 \frac {\partial^2 \tilde H}{\partial p_j \partial q_i} \right) 
 \ee
with $\tilde H$ given by Eq.(\ref{HO2t}).

\subsection{$SO(3)$ gauge oscillator.}

The next in complexity example is the  $SO(3)$ gauge oscillator\cite{Annals}
with the Hamiltonian
 \be
\label{HO3}
\hat H \ =\ \frac 12 \hat P_i^a  \hat P_i^a + 
\frac 12 \omega^2  A_i^a  A_i^a,\ \ \ \ \ \ i,a = 1,2,3 
 \ee
 and the constraints 
 \be
\label{ConO3}
\hat G^a \ =\ \epsilon^{abc} \hat P^b_i A^c_i \ =\ 0\ .
 \ee
$\hat G^a$  can be interpreted as   generators of isotopic gauge 
rotations. 
Only the isosinglet states are present in the physical spectrum.

Proceeding in the same way as above, we can represent the partition
function of this system as the following path integral
 \be
\label{ZO3path}
Z^{O(3)}(\beta) \ =\ \int {\cal D}O(\vecg{\phi})
\int \prod_{ia\tau}  dA_i^a(\tau) \exp\left\{
- \int_0^\beta d\tau \tilde{\cal L}_E[A_i^a(\tau)] \right \}\ ,
 \ee
where
  \be
\label{LEO3}
 \tilde {\cal L}_E[A_i^a(\tau)]\ =\ \frac 12 \left(\dot A_i^a + 
\epsilon^{abc} \frac {\phi^b}\beta A_i^c \right)^2 +
\frac 12 \omega^2  A_i^a  A_i^a \ .
  \ee 
Periodic boundary conditions for $A_i^a(\tau)$ are imposed. Now,
${\cal D}O(\vecg{\phi})$ is the Haar  measure on the $SO(3)$ group
normalized to $\int {\cal D}O(\vecg{\phi}) = 1$. Explicitly,
 \be
\label{measure}
{\cal D}O(\vecg{\phi}) = \frac {d\vecg{\phi} (1 - \cos |\vecg{\phi}|)}
{4\pi^2 \vecg{\phi}^2} \ , \ \ \ \ \
\ \ \ \ \ 0 \leq |\vecg{\phi}| \leq \pi \ .
 \ee
To find the inner integral in Eq.(\ref{ZO3path}), we can set $\phi^a = (0,0,
\phi)$. Then the variables $A_i^3$ are ``neutral'' with respect to $\phi^a$
and 6 remaining ``charged'' variables are decomposed into three pairs, which
are coupled to $\phi$ in the same way as the pair $x_i$ 
was coupled to $\phi$ in the $SO(2)$ case. 

The calculation of the functional integral for the 
partition function involves the products over the modes. Each pair
of charged variables brings about the factor
 \be
\frac 1{(\beta\omega)^2 + \phi^2} \prod_n' 
\frac {\omega_n^2} {(\omega_n + \phi/\beta )^2 + \omega^2}
= \frac 1{2[\cosh(\beta\omega) - \cos \phi]}
 \ee
and each neutral variable provides a $\phi$--independent factor
 \be
\frac 1{\beta\omega} \prod_{n=1}^\infty 
\frac {\omega_n^2} {\omega_n^2 + \omega^2} = 
\frac 1{2 \sinh\left(\frac{\beta\omega}2 \right)}\ .
  \ee
Integrating it over $\vecg{\phi}$ with the measure (\ref{measure}), we obtain the result
 \be
\label{ZO3}
Z^{O(3)}(\beta) \ =\ \frac 1{8 
\sinh^3\left(\frac {\beta\omega}2\right)}
\int_0^\pi \frac {d\phi}\pi (1-\cos \phi) 
\left\{  \frac 1{2[\cosh(\beta\omega) - \cos \phi]} \right\}^3
\nonumber \\
=  \frac 1{64 \sinh^3\left(\frac {\beta\omega}2 \right)}
\frac {2\cosh^2 (\beta\omega) + 1 - 3 \cosh (\beta\omega) }
{2\sinh^5 (\beta\omega)} \ .
 \ee
At large $\beta$,
  \be
\label{ZO3as}
Z^{O(3)}(\beta\omega \gg 1) \ \approx\ 
e^{-9\beta\omega/2} \left[1 + 6 e^{-2\beta\omega} +
e^{-3\beta\omega} + \ldots \right]\ .
   \ee
This asymptotic expansion corresponds to the presence of the ground
state with  energy $9\omega/2$, which is an isosinglet; 
the absence
of isosinglets with energy  $9\omega/2 + \omega$; the presence
of 6 isosinglets with energy $9\omega/2 + 2\omega$ (their
wave functions have the form $\Psi_{ij} \propto A_i^a A_j^a 
\exp\{-\omega (A_k^b)^2/2 \}$; the presence of one isosinglet
with energy $9\omega/2 + 3\omega$ (its wave function involves the
factor $\epsilon_{ijk} \epsilon^{abc} A_i^a A_j^b A_k^c$); etc.

At small $\beta$, 
 \be
\label{ZO3expq}
Z^{O(3)}(\beta\omega \ll 1) \ \approx\ 
\frac 1{32(\beta\omega)^6} \left[1 + \frac {(\beta\omega)^2}8
 + \ldots \right]\ .
   \ee
This result can be reproduced by quasiclassical expansion of 
the integrand in the path integral. 
To leading order, the partition function is given by the integral
 \be
\label{ZO3qint}
 Z^{O(3)}_{\rm quasicl} \ =\  
\int \frac {d\vecg{\phi}}{8\pi^2}
\int \prod_{ia} \frac {dP_i^a dA_i^a}{2\pi} e^{-\beta \tilde H}
  \ee
with
\be
\tilde H \ =\ H + \frac {i\phi^a}\beta G^a \ .
 \ee
If  corrections in quasiclassical expansion are taken into account, 
the integral for the partition function reads
  \be 
 \label{ZO3qint1}
Z^{O(3)}(\beta\omega \gg 1) \ = \ 
\int \frac {d\vecg{\phi}}{8\pi^2} \left(1 - \frac 
{\phi^2}{12} + \ldots \right)
\int \prod_{ia} \frac {dP_i^a dA_i^a}{2\pi} e^{-\beta \tilde H}
\nonumber \\ \left[1 -  
\frac {\beta^2}{24} \left(\frac {\partial^2 \tilde H}{\partial 
P_i^a 
\partial P_j^b}
\frac {\partial^2 \tilde H}{\partial A_i^a \partial A_j^b} - 
 \frac {\partial^2 \tilde H}{\partial P_i^a \partial A_j^b} 
 \frac {\partial^2 \tilde H}{\partial P_j^b \partial A_i^a} \right) 
+ \ldots \right]\ ,
 \ee
where the factor $1 - \phi^2/12$ comes from the expansion of
the factor $2(1-\cos \phi)/\phi^2$ in the measure 
(\ref{measure}) and the correction $\propto \beta^2$ in the inner
integral has the same origin as before. We emphasize that the 
integrals in Eqs.(\ref{ZO3qint}, \ref{ZO3qint1}) are done over the
{\it whole range} of $\vecg{\phi}$.
Calculating the derivatives and performing the  integral over 
$\prod_{ia} dP_i^a dA_i^a$, we obtain
 \be
\label{ZO3bl}
Z^{O(3)}(\beta) \approx \nonumber \\
\frac 1{(\beta\omega)^3} \int_0^\infty \frac {\phi^2 d\phi}
{2\pi} \left(1 - \frac {\phi^2}{12}\right)
 \frac 1{[(\beta\omega)^2 + \phi^2]^3}
\left[ 1 + \frac {2\phi^2 - 3(\beta\omega)^2}{8} \right] \ ,
 \ee
which coincides with the expansion of the integral in 
Eq.(\ref{ZO3})
where the upper limit is set to 
infinity. The integral (\ref{ZO3bl}) converges, but the expansion
up to the terms $\propto  \phi^4,\ \propto  \phi^6$, etc
 results in
  {\it divergent} integrals and only the contribution of the
residues at $\phi = i\omega\beta$ should be taken into account.

The recipe given after Eq.(\ref{expan}) works again. It can be justified
in the same way as above by using the periodicity of the integrand
in the exact expression for $ Z^{O(3)}(\beta)$ in Eq.(\ref{ZO3}).

\section{Supersymmetric quantum mechanics.}
\setcounter{equation}0

  In this section, we will study the fermion--graded
partition function (\ref{ZSUSY}) in the simplest supersymmetric
quantum mechanical system\cite{WitSQM}. The (classical) Hamiltonian of the model is 
 \be
\label{HSQM}
H \ =\ \frac {p^2}2 + \frac 12 [V'(x)]^2 + V''(x) \bar \psi \psi\ ,
  \ee
where $\psi, \bar \psi $ are holomorphic fermion Grassmann
variables. The 
function $V(x)$ is called superpotential. The fermion--graded 
partition function  is given by the Euclidean path
integral where both bosonic and fermionic variables are periodic
in Euclidean time $\tau$. 
\footnote{Recall that the path integral for the ordinary partition
function ${\rm Tr} \{e^{-\beta \hat H}\}$ involves 
antiperiodic boundary conditions for fermionic variables.} 
To leading order, one can 
assume $x,p,\psi$, and $\bar \psi$ to be constant, and we arrive
at the result (\ref{intCec}). Performing the integrals over
$ d\bar\psi d\psi dp$, we obtain
  \be 
  \label{indSQM}
 Z^{F{\rm-grade}} \ =\ \sqrt{\frac \beta{2\pi}} \int_{-\infty}^\infty
dx V''(x) \exp\left\{ - \frac\beta 2  [V'(x)]^2 \right\} \ .
  \ee
If the potential $[V'(x)]^2/2$  grows at large $x$, the spectrum
is discrete and $ Z^{F{\rm-grade}} = \pm 1$ or $ 0$ , depending on the
asymptotics of $V(x)$. $ Z^{F{\rm-grade}}$ defines in this case the
Witten index of the system. In the next--to--leading order, 
$ Z^{F{\rm-grade}}$ is given by the integral 
  \be
 \label{Ceccorr}
Z^{F{\rm-grade}} (\beta) \ =\ \int \prod_i \frac {dp_i dq_i}{2\pi}
\prod_a d\bar\psi^a d\psi^a \exp \left\{ - 
\beta H \right\}(1 + \delta) \ ,
  \ee
 where
   \be
   \label{Delta}
\delta(p_i,q_i; \psi_a, \bar\psi_a) \ =\ \frac{\beta^2}{48}
\left[ \frac {\partial^2}{\partial\Psi_a \partial \bar \psi_a}
-  \frac {\partial^2}{\partial\psi_a \partial \bar \Psi_a}
+ i\left(  \frac {\partial^2}{\partial q_i \partial P_i}
\right. \right. \nonumber \\  
 \left. \left. \left.
-  \frac {\partial^2}{\partial Q_i \partial p_i} \right) \right]^2
H(p_i,q_i; \psi_a, \bar\psi_a) H(P_i,Q_i; \Psi_a, \bar\Psi_a)
\right|_{P=p,Q=q; \Psi = \psi, \bar\Psi = \bar\psi}\ .
  \ee
The formulae (\ref{Ceccorr}), (\ref{Delta}) represent
 a rather straightforward 
generalisation of Eq.(\ref{corrZHpq}) and can be derived either
using the methods of Appendix  [the right side of 
Eq.(\ref{Delta}) is expressed via the star product $H*H$ in
the case when $H$ depends on both bosonic and fermionic 
variables] or with the functional methods of Sect.2. We leave it
to the reader as an exercise.

For the Hamiltonian (\ref{HSQM}) under consideration,
  \be
 \label{DelSQM}
\delta \ =\ - \frac{\beta^2}{24}[V'(x) V^{(3)}(x) + V^{(4)}(x) 
\bar\psi \psi ]\ .
  \ee
Substituting it in Eq.(\ref{Ceccorr}) and doing the integral over
$d\bar\psi d\psi dp$, we obtain
  \be
 \label{popind}
\Delta Z^{F{\rm-grade}} (\beta)\  =\ 
\frac{\beta^{3/2}}{24\sqrt{2\pi}} \int_{-\infty}^\infty dx
[V^{(4)} - \beta V' V'' V^{(3)} ] \exp \left\{- \frac
{\beta [V']^2}2 \right\}  \nonumber \\
 = \ \frac{\beta^{3/2}}{24\sqrt{2\pi}} \int_{-\infty}^\infty dx
\frac {\partial}{\partial x}   
\left[ V^{(3)} \exp \left\{- \frac
{\beta [V']^2}2 \right\} \right] = 0\ .
  \ee
In other words, the corrections to the leading order result for
$Z^{F{\rm-grade}} (\beta)$ vanish in this case, as should have been
expected in advance.
 
 Actually, the correction (\ref{popind}) {\it always} vanishes, even for the 
systems with continuum spectrum, where 
${\rm Tr}\{(-1)^F e^{-\beta \hat H} \}$ may 
depend on $\beta$. As an example, consider the 
Hamiltonian (\ref{HSQM}) with 
  \be
   \label{Vlnch} 
V(x) \ =\ \ln \left[ \cosh\left(\frac xa \right) \right]\ .
   \ee
  In this case, $V'(x) = (1/a) \tanh (x/a)$ tends to a constant at large $x$ 
and the spectrum is continuous. The fermion--graded partition function depends 
on $\beta$ and it is seen already in the leading quasiclassical order,
  \be
    \label{Zlnch}
  Z^{F{\rm-grade}} (\beta) \ =\ \Phi\left( \frac {\sqrt{\beta}}a \right)\ ,
 \ee
 where $\Phi(x)$ is the probability integral. This expression alone provides 
the correct asymptotics  $\lim_{\beta \to \infty}  Z^{F{\rm-grade}} (\beta)  = 1$
 corresponding to the presence of the normalized vacuum state with
  \be
  \label{vaclnch}
\Psi(x) \propto \ e^{-V(x)} = \ \frac 1{\cosh\left(\frac xa \right)} \ .
   \ee
  And the integral (\ref{popind}) and probably also all 
higher--order terms in the quasiclassical expansion  vanish.

As another example with continuous spectrum and mismatch between
 ${\rm Tr }\{(-1)^F e^{-\beta \hat H}\}$ and the 
Witten index, consider the 
superconformal SQM with $V(x) = \lambda \ln(x/a)$ \cite{SCQM}. 
If $\lambda
> 1$, we have the singular repulsive potential $\propto 1/x^2$
in both the bosonic and fermion sectors so that the motion is
restricted to the half--line $x \in (0,\infty)$. Hence, the integral
(\ref{indSQM}) is done within the limits $0 \leq x < \infty$, and we obtain
$ Z^{F{\rm-grade}} (\beta) = 1/2$ in the leading order.
The integral 
(\ref{popind}) and all higher--loop corrections vanish, however.
In this case, it is not so surprising. The Hamiltonian (\ref{HSQM})
with $V(x) \propto \ln x$ does not involve a dimensionful
parameter and neither $E_{\rm char}$ nor the dimensionless
parameter of the quasiclassical expansion $\beta E_{\rm char}$
can be defined.
  
\section{SYM quantum mechanics.}
\setcounter{equation}0

There are systems, however, 
for which the quasiclassical series for $ Z^{F{\rm-grade}} (\beta) $ is 
expected to be ``alive''. In this section we consider the gauge 
supersymmetric quantum 
mechanical systems obtained by dimensional reduction
from  ${\cal N} =1$ SYM field 
theories in 4, 6, 
and 10 dimensions. We will refer to them as ${\cal N} =2$ , 
${\cal N} =4$, and ${\cal N} =8$ SYM QM systems, indicating the number 
of different complex supercharges $Q_\alpha$.
Consider first the
${\cal N} =2$ system  described by the 
Hamiltonian (\ref{hamSYM}) and the constraints (\ref{conSYM}). 
 To leading order, the partition function is given by the 
integral \cite{jaIW,Sethi} 
   \be
\label{intjaSeth}
   Z^{\rm SYM}({\rm small} 
\ \beta) = \frac 1{V_G} \int \prod_a \ d(\beta g A_0^a)  
  \int \prod_{ai} \frac {dP_i^a dA_i^a}{2\pi} \prod_{\alpha a} 
d\bar\lambda^{a\alpha} d\lambda_{\alpha}^a \nonumber \\
\exp\{-\beta[H + ig \vec{A}_0 \vec{G}] \} \ ,
 \ee
 where $ \beta g A_0^a \equiv \phi^a$ are gauge parameters and $V_G$
is the volume of the gauge group. The notation $\vec{A}_0 \vec{G}
\equiv A_0^a G^a$ is used. 

As was mentioned in the introduction, the calculation of the 
integral (\ref{intjaSeth}) gives a fractional number, while
the Witten index $Z^{\rm SYM}(\beta = \infty)$ is integer
(it is zero for the ${\cal N} =2$ and ${\cal N} =4$ systems).
Also, the Hamiltonian (\ref{hamSYM}), in contrast to the 
Hamiltonian of the superconformal quantum mechanics considered
at the end of the previous section, involves a dimensional 
parameter $g $ [remember, we are not in (3+1), but in
(0+1) dimensions !]. The characteristic energy scale is 
$E_{\rm char} \sim g^{2/3}$ (this estimate is easily
obtained by equating the characteristic kinetic and 
potential energies,
$1/A^2 \sim g^2A^4$), and we expect $Z^{\rm SYM}(\beta)$
to be a nontrivial function of $\beta$ tending to the limit
(\ref{intjaSeth}) when $\beta g^{2/3} \ll 1$ and to the integer
Witten index when  $\beta g^{2/3} \gg 1$.       

 We want to calculate the correction $\propto \beta^2$ to the 
result (\ref{intjaSeth}). Using the experience acquired when 
studying toy models 
in the previous sections, we understand that we have to {\it (i)}
  take into account the expansion of the Haar measure on $G$ 
and  {\it (ii)}
calculate the correction $1 + \delta$ in the integrand, where
   $\delta(P^a_i, A^a_i; \lambda_\alpha^a, \bar \lambda^{a\alpha})$
   is evaluated using the rule (\ref{Delta}) with
    \be
    \label{HtoHt}
    H \ \longrightarrow \ \tilde H = H + ig\vec{A}_0 \vec{G} \ .
      \ee
Let us first evaluate the contribution of the bosonic derivatives
in Eq.(\ref{Delta}). 
      The direct calculation gives
       \be
       \label{delbos}
       \delta_{\rm bos}^{{\cal N}=2}
 \ =\ - \frac {\beta^2 g^2 c_V \vec{A}_i^2}{12}  +
 \frac {\beta^2 g^2 c_V \vec{A}_0^2}8\ ,
 \ee
 where 
$c_V$ is the adjoint Casimir eigenvalue. The first term in the
right side of Eq.(\ref{delbos}) comes from differentiation of $H$
and the second term from differentiation of $ig\vec{A}_0 \vec{G}$
in Eq.(\ref{HtoHt}). 

One can also easily perform the calculation for the ${\cal N}=4$
and ${\cal N}=8$ systems. The bosonic part of the Hamiltonian and
constraints 
have the same form as in Eqs.(\ref{hamSYM}), (\ref{conSYM}), 
only the spatial index
runs now from 1 to $D-1 = 5,9$. We obtain
  \be
\label{delbosD}
  \delta_{\rm bos}
 \ =\ \frac  {\beta^2 g^2 c_V }{24} \left[ - (D-2) \vec{A}_i^2  +
(D-1) \vec{A}_0^2 \right]\  .
 \ee
In the ${\cal N} = 2$ case the contribution of the fermion derivatives 
in Eq.(\ref{Delta}) is
  \be
 \label{delferm}
 \delta_{\rm ferm}^{{\cal N}=2}
 \ =\  \frac {\beta^2 g^2 c_V}{12} ( \vec{A}_i^2 -
  \vec{A}_0^2 )\ .
 \ee
Consider now  ${\cal N} = 4$ theory.
The fermion terms in the 
Hamiltonian and constraints have, again, the same
form as Eqs.(\ref{hamSYM}), (\ref{conSYM}), but 
$\sigma_i$ are replaced by 5--dimensional Euclidean 
$\gamma$ matrices, $\gamma_i \gamma_j + \gamma_j \gamma_i
= \delta_{ij},\ i,j = 1,\ldots,5$,   
and the spinorial index $\alpha$ runs now from 1 to 4 
(the variables $\lambda_\alpha^a$ are a result of reduction of  
6--dimensional Weyl spinors). Twice as much spinorial components
bring about the factor 2 in Eq.(\ref{delferm}). 

For the  ${\cal N} = 8$ system obtained by dimensional reduction
from 10--dimensional SYM theory, we meet a difficulty: the $SO(9)$ group
admits only real spinor representations and, though one can still
introduce complex holomorphic fermion variables, the Hamiltonian
expressed in these terms does not have a natural structure, the fermion
number is not conserved, etc. It is more convenient to write the
fermionic term in $\tilde H$ as
\be
\label{HfermN8}
\tilde H_{\rm ferm}^{{\cal N} = 8} 
\ =\ \frac {ig}2 f^{abc} \lambda_\alpha^a
(\Gamma_\mu)_{\alpha\beta} \lambda_\beta^b A_\mu^c \ ,
 \ee
where $ \lambda_\alpha^a$ are now real fermion variables; 
$\alpha,\beta = 1,\ldots,16$ and 
$\mu = 0,1\ldots,9$. 
When $\mu = j$ is spatial, $\Gamma_j$ are $16 \times 16$ 
 real and symmetric  9--dimensional $\Gamma$ matrices,
$\{\Gamma_j, \Gamma_k\}_+ = 2\delta_{jk}$. Also, $\Gamma_0 = i$.
Eight holomorphic variables in terms of which Eq.(\ref{Delta}) is
written are expressed via $\lambda_\alpha^a$ as 
 \be
\label{muvialam}
\mu_1^a = \frac 1{\sqrt{2}} (\lambda_1^a + i\lambda_9^a),\ \ 
\ldots \ \ ,\ 
 \mu_8^a = \frac 1{\sqrt{2}} (\lambda_8^a + i\lambda_{16}^a)\ .
 \ee
If using the variables $\lambda$ instead of $\mu$, the fermionic
contribution to $\delta$ is cast in the form
 \be
\label{delfermlam}
 \delta_{\rm ferm}^{{\cal N}=8}  = 
\left. - \frac {\beta^2 g^2}{192} f^{abc} f^{def} A_\mu^a A_\nu^d
\left(  \frac {\partial^2}{\partial\lambda_\alpha^g \partial 
\Lambda_\alpha^g} \right)^2 
 \lambda^b \Gamma_\mu \lambda^c  \Lambda^e \Gamma_\nu \Lambda^f
\right|_{\Lambda = \lambda} \nonumber \\
=   \frac {\beta^2 g^2 c_V}{3} ( \vec{A}_i^2 -
  \vec{A}_0^2 ) \ .
  \ee
 Combining this with Eq.(\ref{delferm}) and with the twice as large
value of $\delta_{\rm ferm}$ for ${\cal N} = 4$, we can write
   \be
 \label{delfermD}
 \delta_{\rm ferm}
 \ =\  \frac {\beta^2 g^2 c_V(D-2) }{24} ( \vec{A}_i^2 -
  \vec{A}_0^2 )\ .
 \ee
Adding Eqs.(\ref{delfermD}) and (\ref{delbosD}) 
(the mixed contributions involving both bosonic and fermionic
derivatives vanish in this case), we obtain
 \be
 \label{delbezmerD}
 \delta
 \ =\  \frac {\beta^2 g^2 c_V}{24} 
  \vec{A}_0^2 \ .
 \ee

This is not yet the full story. The total correction is obtained
if adding to Eq.(\ref{delbezmerD}) the correction coming from the
expansion of the measure. We have already calculated this correction
for the $SU(2)$ group. The corresponding factor in the measure is
  \be
  \label{measSYM}
  \frac {2[1 - \cos (\beta g|\vec{A}_0|) ]}{\beta^2 
 g^2 \vec{A}_0^2} 
\ =\ 
  1 - \frac {\beta^2 g^2 \vec{A}_0^2}{12} + \ldots 
   \ee
We see that the correction coming from the measure
 cancels exactly the correction (\ref{delbezmerD}) and the total
correction $\propto \beta^2$ to the integrand in 
Eq.(\ref{intjaSeth}) vanishes !

One can show that this cancellation works not only for $SU(2)$, but
for any gauge group $G$.   
The measure on an arbitrary Lie group is given by the Weyl formula.
To write it, represent an element $g$ of the group $G$ as
$$ g = \ \Omega^{-1} e^{ih} \Omega\ ,$$
where $h$ belongs to the Cartan subalgebra ${\mathfrak h}$ of the
corresponding  Lie algebra ${\mathfrak g}$ and $\Omega$
belongs to the coset $G/T$, where $T = [U(1)]^r$ is the 
{\it maximal torus} in $G$, $r$ is the rank of the group.
After integrating over angular variables (residing in $\Omega$),
we obtain (see e.g. \cite{group})
  \be
\label{measWeyl}
d\mu_G \ \propto\ \prod_i \sin^2 \left( \frac {\alpha_i(h)}2 
\right)\ ,
  \ee
where the product runs over all positive {\it roots} $\alpha_i$ 
of ${\mathfrak g}$.

We remind that the roots are certain linear forms on 
${\mathfrak h}$. Each positive root $\alpha$ correspond to a pair of 
root vectors
$e_\alpha, e_{-\alpha}$ such that
$$[h, e_{\pm \alpha} ] = \ \pm \alpha(h) e_{\pm \alpha} $$
for any $h \in {\mathfrak h}$. The expansion of the measure
(\ref{measWeyl}) at small $h$ gives
  \be
d\mu_G \propto  \prod_i \alpha_i^2(h) 
\left[ 1 - \frac 1{12} \sum_i\alpha_i^2(h) \right]\ .
 \ee
Using the identity\cite{group}
 \be
\label{Trh2}
 \sum_i\alpha_i^2(h) \ =\ c_V {\rm Tr} \{ h^2 \}\ ,
  \ee
where $h$ are the matrices in the fundamental representation,
substituting
$h = \beta g A_0^a t^a$, and restoring the angular variables, 
we obtain
   \be
\label{measG}
d\mu_G \propto  \prod_{a=1}^{{\rm dim}(G)}
dA_0^a 
\left( 1 - \frac {\beta^2 g^2 c_V}{24} 
\vec{A}_0^2 \right)\ ,
 \ee    
which cancels with Eq.(\ref{delbezmerD}).

Let us illustrate this general result by calculating the correction
to the measure for the group $SU(3)$. 
After conjugating $A_0^a t^a$ to the Cartan subalgebra, we 
can write 
$$h = \frac {\beta g}2 \ {\rm diag}\left(A_0^3 + \frac {A_0^8}{\sqrt{3}}, \ 
 -A_0^3 + \frac {A_0^8}{\sqrt{3}},\  -2  
\frac {A_0^8}{\sqrt{3}} \right)\ .$$
There are three positive roots: 
$$\alpha_1 = (1,-1,0),\ \  \alpha_2 = (0,1,-1)\ \  {\rm and}\ \ 
\alpha_3 = \alpha_1 + \alpha_2 = (1,0,-1)\ .$$
The measure (\ref{measWeyl}) has the form
   \be
\label{measSU3}
d\mu_{SU(3)} \ \propto\  \nonumber \\
\sin^2 \left( \frac {\beta g 
A_0^3}2 \right)
\sin^2 \left( \frac {\beta g (-A_0^3 + A_0^8 \sqrt{3})}4 \right)
\sin^2 \left( \frac {\beta g (A_0^3 + A_0^8 \sqrt{3})}4 \right)\ .
  \ee
Expanding this, we obtain the factor
$$1 - \frac {\beta^2}8 \left[ (A_0^3)^2 +  (A_0^8)^2 \right] $$
 in agreement with (\ref{measG}).

Thus, we arrive at a rather unexpected and remarkable result:
the total correction $\propto \beta^2$ to the fermion--graded
partition function of the SYM quantum mechanics vanishes for 
all groups and  all ${\cal N}$.

\section{${\cal N} =2$ SUSY $\sigma$ model on $S^3$.}
\setcounter{equation}0

Consider the system (\ref{QHN2}) with
\be
\label{fS3}
f(\vec{x}) = 1 + \frac {\vec{x}^2}{4R^2} \ .
  \ee
The bosonic part of the Hamiltonian coincides up to a  constant 
shift 
with the Laplacian on $S^3$,
 \be
\label{HbosS3}
\hat H_{\rm bos} \ =\ - \frac 12 \Delta_{S^3} + \frac 3{8R^2}\ ,
  \ee
where the metric is written in the stereographic coordinates,
 \be
\label{metS3}
ds^2 \ =\ \frac {d\vec{x}^2}{f^2} = \frac {d\vec{x}^2}{\left( 1 
 + \frac {\vec{x}^2}{4R^2}\right)^2} \ . 
  \ee
The relation
$$r = \ |\vec{x}| \ =\ 2R\tan \frac \Theta 2$$
holds, where $\Theta$ is the polar angle on $S^3$. The constant $R$ is
the radius of the sphere. The Hamiltonian (\ref{HbosS3}) acts upon the wave
functions with canonical normalization
\be
\label{normS3} 
 \int  |\Psi |^2 d\vec{x} \ =\ 1\ .
 \ee
Alternatively, one can perform a unitary transformation and define
$\hat H_{\rm cov} = f^{3/2} \hat H f^{-3/2}$, which acts on the 
covariantly normalized
wave functions $\Psi_{\rm cov} = f^{3/2} \Psi$,
 \be
\label{normcov}
\int   |\Psi_{\rm cov} |^2   \sqrt{g} d\vec{x} = 
\int   |\Psi_{\rm cov} |^2   \frac {d\vec{x}}{f^3} = 1\ .
 \ee
Now, $S^3$ is a compact manifold, the motion is finite, and the spectrum
is discrete. There are two bosonic supersymmetric vacua, which are annihilated
upon the action of the Hamiltonian and supercharges (\ref{QHN2}). Their wave
functions are
 \be
\label{volnfun}
\Psi_{\rm vac}^{(1)} \propto f^{-1} (\vec{x}),\ \ \ \ \ 
\Psi_{\rm vac}^{(2)} \propto \psi^2 f^{-1} (\vec{x}) \ .
  \ee
The functions (\ref{volnfun}) correspond to the covariant wave functions
 \be
\label{psicov}
\Psi_{\rm cov} \propto \sqrt{f} = \sqrt{  1 + \frac {\vec{x}^2}{4R^2} }
= \frac 1{\cos \frac \Theta 2} \ .
  \ee
The wave function (\ref{psicov}) is singular at the north pole of the
sphere, but this singularity is of a benign, normalizable kind.

Let us discuss this important point in some details. When one considers a
purely mathematical problem of the spectrum of the Laplacian on a sphere, only
nonsingular eigenfunctions are usually considered. One of the reasons for
that is that the function (\ref{psicov}) is not strictly speaking an
eigenfunction of the Laplacian, the action of $\Delta_{\rm cov}$
on (\ref{psicov}) gives besides $(-3/8R^2) \Psi_{\rm cov}$ also a
$\delta$ function   singularity at $\Theta = \pi$. However, if one considers
$S^3$ with its north pole removed, nothing prevents us to bring into 
consideration singular normalizable wave functions of the type 
(\ref{psicov}).

Consider now the standard ${\cal N} =1$ supersymmetric quantum mechanics on
$S^3$ \cite{witten}. The bosonic part  of the Hamiltonian is, again, 
the Laplacian, but in this case the proper Hilbert space does {\it not}
include singular wave functions. The matter is that, though the function
(\ref{psicov}) is normalizable, the function obtained from it by the action
of the ${\cal N} =1$ supercharge $\hat Q$ (we remind that 
$\hat Q^{{\cal N} =1}$ is the operator
of external differentiation acting on the forms) is not.
Indeed,  
 $$\hat Q^{{\cal N} =1} f^{1/2}  
\sim e^i_a \psi^a \partial_i f^{1/2}
\propto  f\partial_i 
f^{1/2}\  \propto r^2 $$
(the vielbein $e^i_a$ was chosen in the form 
$e^i_a = f\delta_{ai}$),
and the normalization integral (\ref{normcov}) 
diverges linearly at large distances. Therefore, the function (\ref{psicov})
does not have a normalizable 
superpartner and is not admissible by that reason\cite{SSV}.

In our case, however, the wave functions (\ref{volnfun})
are annihilated by the supercharges $\hat Q_\alpha, 
\ \hat{\bar Q}^\alpha$, and there is no reason whatsoever to 
ignore them. In the sector with $F=1$, $\Psi(\vec{x}, \psi_\alpha)
= P^\alpha(\vec{x})\psi_\alpha$, normalizable solutions to the 
equations $\hat Q_\alpha \Psi = 
\hat {\bar Q}^\alpha \Psi = 0$ are absent. Thus, there are no
normalizable fermionic vacuum states and the Witten index is
equal to $2-0 = 2$. 

Let us calculate now the functional integral for the fermion--graded
partition function. As the spectrum is
discrete,  one could expect {\it a priori} that, as it was the
case for the simple  models of Sect. 4, the leading
order calculation gives the correct result ${\rm Tr}\{(-1)^F
e^{-\beta \hat H} \} = 2$, and the higher--order corrections vanish.
The actual situation is much more interesting.

To leading order, the  fermion--graded
partition function is given by the integral (\ref{intCec}). 
A novelty is that the value of this integral
 depends in essential way on how the ordering
ambiguities are resolved and
 the classical Hamiltonian is chosen. One of
the choices is to define $H^{\rm cl}$ as the Poisson bracket 
of two classical supercharges, which are defined in turn as the Weyl
symbols of the quantum ones \cite{quantiz} and have the same
functional form as the expressions (\ref{QHN2}) up to the change
$\hat p_k \to p_k,\ \ \hat{\bar \psi}^\alpha \to \bar 
\psi^\alpha$. In this case, 
 \be
\label{Hclryba}
 H^{\rm cl}  = 
\frac 12\{Q_\alpha, \bar Q^\alpha\}_{P.B.} =  \nonumber \\
\frac 12 f^2(\vec{x})   p_k^2 
 - \epsilon_{jkp}  {\bar\psi} \sigma_j \psi  f(\vec{x})
\partial_p f(\vec{x})  p_k
- \frac 12  f(\vec{x}) \partial^2_k f(\vec{x})
( \bar\psi \psi )^2\ . 
  \ee
Substituting this in Eq.(\ref{intCec}) and performing the 
integral over momenta and fermion variables,
 we obtain 
 \be
\label{total}
Z^{F{\rm -grade}}\  = \frac 1{\sqrt{\beta} (2\pi)^{3/2}}
\int d\vec{x}\ \partial_k \left( \frac {\partial_k f}{f^2}
\right) \ =\ 0\ .
 \ee
Note the presence of the large factor $\propto 1/\sqrt{\beta}$ in
front of the integral. The latter vanishes, however.
 
 Another option is to use $\tilde H^{\rm cl}$ 
 defined as the Weyl symbol of the
quantum Hamiltonian in Eq.(\ref{QHN2}). Weyl symbol of an 
anticommutator 
$\{\hat Q_\alpha, \hat {\bar Q}^\alpha\}_+$ 
is not given by the Poisson bracket of the Weyl
symbols $ Q_\alpha^{\rm cl}, {\bar Q}^{\alpha\,{\rm cl}}$, but rather
by their {\it Moyal bracket}.
\footnote{We remind its definition in the Appendix.} 
 We derive
 \be
\label{HMoybra}
 \tilde   H^{\rm cl}  =  H^{\rm cl} + \frac 14 [\partial_k 
f(\vec{x})]^2 \ . 
  \ee
The fermion--graded partition function is
  \be
\label{leadord}
Z^{F{\rm -grade}}\ = \frac 1{\sqrt{\beta} (2\pi)^{3/2}}
\int d\vec{x} \ \partial_k \left( \frac {\partial_k f}{f^2} \right)
\exp\left\{-  \frac \beta 4 [\partial_k 
f(\vec{x})]^2 \right\} \nonumber \\
 \ =\ 2\sqrt{2} + o(\beta) 
 \ee
for $f(\vec{x}) = 1 + \vec{x}^2/(4R^2)$. Neither 0 nor $2\sqrt{2}$
is the correct result for the index, which means that, 
contrary to naive expectations, higher--order terms in the 
quasiclassical expansion must be relevant in this case. 

And they are. 
First, note that, for small $\beta$, the characteristic values
of $r = |\vec{x}|$ are large: $r_{\rm char}
\sim R^2/\sqrt{\beta}$. Also, one can estimate 
$$ \beta H^{\rm cl} \sim 1\ \longrightarrow \ 
\beta \frac{r^4}{R^4} p^2 \sim 1  \ \longrightarrow 
\ p_{\rm char} \sim
\frac {R^2} {r_{\rm char}^2 \sqrt{\beta}} \sim \frac{\sqrt{\beta}}
{R^2}\ . $$
It follows that $p_{\rm char} r_{\rm char} \sim 1$ and
the correction $\delta$ 
 calculated according to the rule (\ref{Delta}) is estimated
to be of order 1. Thus, the ``correction $\propto \beta^2$\ ''
does not depend on $\beta$ at all in this case, but is simply
a number !
The same concerns the higher--loop corrections: they are 
all equal
to some numbers, and there is no expansion parameter.

This remarkable result can be given the following explanation.
First, for a supersymmetric system with discrete spectrum, 
${\rm Tr}\{(-1)^F
e^{-\beta \hat H}\}$ just cannot depend on $\beta$ and
that refers also to the individual terms in the quasiclassical 
expansion. If higher--loop corrections appear, they have to be
$\beta$--independent numbers.  
Second, as was discussed before, the proper quasiclassical expansion
parameter is $\beta E_{\rm char}$. For the system under 
consideration, the charateristic
energy is determined by the radius of the sphere:
 $ E_{\rm char} \sim 1/R^2$.   
On the other hand, $r_{\rm char} \propto \beta^{-1/2}$ are large,
which means that the integral is saturated by the region at the
vicinity of the north pole, where the metric is almost
flat. In other words, our integral does not
``know'' about the existence of the sphere and about the value
of $R$. It does not depend  on $E_{\rm char}$ and cannot thereby
depend on $\beta$.
The situation is similar 
to the situation for the superconformal quantum mechanics, 
discussed at the end of Sect. 4. The difference is that in the
case of superconformal quantum mechanics higher--order 
corrections vanish. Here, all such corrections have the same order and
the quasiclassical expansion breaks down.
 
 \section{Discussion.}
Ê  \setcounter{equation}0

The main subject of this paper was the studying of the quasiclassical 
expansion of  $Z^{F{\rm-grade}} (\beta)$ for supersymmetric systems. 
But the results obtained in Sect. 2,3 for purely bosonic systems also
 present a certain methodical interest. 

To derive the expression (\ref{corrZHpq}) by functional methods, we had to 
solve the spectral problem with generic quadratic Hamiltonian (\ref{HPauli}). 
The explicit solution is given in Eqs.(\ref{spectr}), (\ref{detOm}).
\footnote{We were not able to find these results in the literature.}

An important particular case is the Hamiltonian describing multidimensional
motion in a constant magnetic field $F_{ij}$ with generic oscillatoric 
potential,
 \be
  \label{HamFS}
  \hat H \ =\ \frac 12 \left( \hat p_i - \frac 12 F_{ij} x_j \right)^2
  + \frac 12 S_{ij} x_i x_j \ .
  \ee
The eigenvalues $\Omega_j$ are determined by  the roots $\lambda^{(j)}$ of 
the polynomial
 \be
 \label{polFS}
 \det \|\lambda + \sqrt{\lambda}F + S \|\ ,
 \ee
$\lambda^{(j)} = - \Omega_j^2$.
 If all eigenvalues of $S$ are positive, the quadratic form $S_{ij} x_i x_j$
 is positive definite and the whole differential operator (\ref{HamFS}) is positive
 definite. Hence, it has a real positive spectrum (\ref{spectr}) and hence all 
$\Omega_j$ and $\Omega_j^2$ are real and positive.

We have derived a purely mathematical result: for a positive definite $S$ and
antisymmetric $F$,
all  roots of the polynomial (\ref{polFS}) 
are real and negative. This simple but amusing fact can also be derived by
purely algebraic means \cite{Rehren}. As $A =  \lambda + \sqrt{\lambda}F + S$
has zero determinant, it has an eigenvector $v$ with zero eigenvalue,
$Av = 0$ and hence $\langle v, Av \rangle = 0$. We obtain
$\lambda \langle v, v\rangle + \langle v, Sv \rangle = 0$\ \ 
($\langle v, Fv \rangle = 0$ due to antisymmetry of $F$). As $S$ is positive
definite, $\lambda$ must be real and negative.

In Sect. 3 we studied the quasiclassical expansion of $Z$ for gauge QM
systems. We have learned  that,  when calculating the corrections, 
{\it (i)} We have to extend the limits of integration over the gauge 
parameters to infinity even if the gauge group is compact.
{\it (ii)} If the integral thus obtained diverges, the correction is still 
finite and is given by the residue of the integrand at the same pole which shows
up in the integral to leading order. 

We believe that this lesson might help to justify the calculation of the leading 
quasiclassical contribution to  $Z^{F{\rm-grade}} (\beta)$
in SYM quantum mechanics, performed in Refs.\cite{Nekr,Stau}. (To calculate the integral, 
the authors of Refs.\cite{Nekr,Stau} first
deformed the system by introducing mass parameters and then replaced the 
divergent integrals by the pole contributions as if the integrals were 
convergent.)

Our initial guesses when studying supersymmetric QM systems were that
 \begin{enumerate}
 \item For the systems with discrete spectrum, $Z^{F{\rm-grade}}$, which does not 
depend on $\beta$, is calculated correctly in the leading quasiclassical approximation
and all corrections vanish. 
 \item For the systems with continuous spectrum, the quasiclassical series 
comes to life. The sum of the series gives a nontrivial function
$Z^{F{\rm-grade}} (\beta)$ determining the index (\ref{IWit}) in the
limit $\beta \to \infty$. 
 \end{enumerate}
 
 Unexpectedly, we have discovered a lot of other scenarios.
  \begin{enumerate}
 \item For 1--dimensional SQM systems with continuous spectrum, the function is 
determined by the leading quasiclassical contribution as it is the case when 
the spectrum is discrete. (The difference is that, for the system with continuum 
spectrum, $Z^{F{\rm-grade}}$ may depend on $\beta$.)
The loop corrections vanish.
\footnote{We have checked it only at the 1--loop level, but our conjecture 
is that the corrections vanish also for higher loops. Indeed, higher-loop 
corrections must vanish for the systems (\ref{HSQM}) with
discrete spectrum. This implies that the corresponding contribution is 
reduced to the integral of total derivative, as it was the case for
the 1--loop corrections. But then the integral must vanish for any superpotential 
$V(x)$. }
 \item  The 1--loop corrections vanish also for the SYM systems 
for all ${\cal N} = 2,4,8$ and  for all groups. (The cancellation occurs if 
all the effects of order $\beta^2$, including  the expansion of the Haar 
measure, are taken into account.) 

Actually, this result may be not so surprising.
\footnote{The reasoning below belongs to A. Vainshtein.}
Indeed,  guess 2 of the previous list was based mainly on the known 
calculation for the index where the trace was regularized by putting a boundary
in the field space, which makes the spectrum discrete. The associated
boundary conditions break supersymmetry and result in $\beta$-dependence
of the fermion-graded partition function. In this paper, we adopted a different
philosophy. We {\it defined}  $Z^{F{\rm-grade}}(\beta)$ via the corresponding
path integral and studied the quasiclassical expansion for the latter. It is
reasonable to expect that the path integral still does not depend on $\beta$
if the characteristic values of the fields contributing to the 
integral are not large so that, even if the boundaries are set, their effect
[breaking of supersymmetry and $\beta$-dependence of  
$Z^{F{\rm-grade}}(\beta)$] is not felt yet. This is so for the leading order 
integral (\ref{intjaSeth}): it converges with 
$A^{\rm char} \sim (\beta g^2 )^{-1/4}$.
The integrals with account of ``individual'' corrections of 
Eq.(\ref{delbos}) etc
converge in a similar way for ${\cal N} = 4$ and  ${\cal N} = 8$ theories.
There is a potential logarithmic divergence in the  ${\cal N} = 2$ case,
but seemingly this divergence is not strong enough to make the fermion-graded
partition function $\beta$--dependent. 

At the two--loop and higher level, the integrals start to diverge
as a power. Though the answer obtained using the recipe of Sect. 3 should
be finite, we do not expect the coefficient of $\beta^4$ in the 
quasiclassical expansion of $Z^{F{\rm-grade}}(\beta)$ to vanish.
 An explicit calculation of such 
corrections is an interesting though not so simple problem.
\item On the other hand, we have found a system with discrete spectrum, 
described in Eq.(\ref{QHN2}), where the corrections to the leading 
quasiclassical result do not vanish, though they do not depend on $\beta$ 
and are all of the same order. 

A following interpretation of this fact can be suggested. In the standard 
supersymmetric $\sigma$--model on a compact manifold, the Witten index 
coincides with the Euler characteristics $\chi$ of the manifold. The leading 
term in the quasiclassical expansion for
$Z^{F{\rm-grade}}$ is none other than the known integral representation for 
$\chi$. In the nonstandard model (\ref{QHN2}), the
Witten index is equal to 2 for $S^3$, while $\chi(S^3) = 0$. One can
represent $2 = \beta_0 + \beta_3$, where $\beta_i$ are the 
Betti numbers, but 
no integral representation for this quantity is known.
And that is why a ``normal" scenario for a system with discrete spectrum --- 
$Z^{F{\rm-grade}}$  is determined by the leading--order
formula and the corrections vanish --- is not realized in this case.  
 \item We did {\it not} find a system where the 1--loop correction to
$Z^{F{\rm-grade}} (\beta)$ would have a normal order 
$\sim (\beta E_{\rm char})^2$ 
with a nonvanishing coefficient. It would be interesting to find one.  
 \end{enumerate}

 I am indebted to K.-H. Rehren, D. Robert and A. Vainshtein for 
illuminating discussions.

 \section*{Appendix: Quasiclassical expansion and star product.}
 \setcounter{equation}0
 \renewcommand{\theequation}{A.\arabic{equation}}

 We outline here how the result (\ref{corrZHpq}) for the quasiclassical 
correction to the partition function of a quantum system is derived using 
operator methods. Bearing in mind the representation (\ref{ZexpWeyl}), 
the problem is reduced to  evaluating 
 the Weyl symbol of the exponential $\exp\{-\beta \hat H \}$. The Weyl symbol
of the product of two operators is given by the expression \cite{Weyl}
  \be
  \label{starprod}
  [\hat A \hat B]_W  = \left. \exp \left\{ \frac {i\hbar}2 
  \left( \frac {\partial^2}{\partial q_i \partial P_i} - 
  \frac {\partial^2}{\partial p_i \partial Q_i} \right) \right\}
  A(p_i, q_i) B(P_i, Q_i) \right|_{P=p,Q=q}\ ,
  \ee
  where $A(p_i,q_i)$ and $B(P_i,Q_i)$ are Weyl symbols of the operators 
$\hat A, \hat B$. Now, $\hbar$ is the Planck's constant, which we preferred 
to retain here to facilitate bookkeeping.

In the modern language, the right side of Eq.(\ref{starprod}) is called 
the {\it star product} $A*B$ of the functions  $A(p_i,q_i)$ and $B(p_i,q_i)$ 
\cite{noncom}. The star product is not commutative. The star commutator 
$A*B - B*A$ is called the {\it Moyal bracket}
of the functions $A,B$. In the leading order in  $\hbar$, the Moyal bracket 
is reduced to the Poisson bracket. The star product is associative, 
however: $(A*B)*C = A*(B*C)$.

We need to determine 
  \be
  \label{WexpH}
  \left[ e^{ - \beta \hat H} \right]_W  = 
  1 - \beta H + \frac {\beta^2}2 H*H - \frac {\beta^3}6 H*H*H + \ldots\ \ \ , 
  \ee 
where $H$ is the Weyl symbol of the Hamiltonian. Keeping only the
 terms $\propto 1$ and $\propto \hbar^2$, we obtain
 \be
 \label{HstarH}
 H*H = H^2 - \frac{\hbar^2}4 \left( \frac {\partial^2 H}{\partial p_i 
\partial p_j } \frac {\partial^2 H}{\partial q_i \partial q_j }
- \frac {\partial^2 H}{\partial p_i \partial q_j } \frac {\partial^2 H}
{\partial p_j \partial q_i} \right) \nonumber \\
\stackrel{\rm def}=  \ H^2 + \hbar^2 \Delta + o(\hbar^2)\ .
  \ee
Further,
   \be
 \label{H3star}
 H*H*H = H^3 + 3 \hbar^2 \Delta  H - \frac{\hbar^2}4 \left( \frac 
{\partial^2 H}{\partial q_i \partial q_j } \frac {\partial H}{\partial p_i }
\frac {\partial H}{\partial p_j }
 -2  \frac {\partial^2 H}{\partial p_i \partial q_j } \frac 
{\partial H}{\partial p_j } \frac {\partial H}{\partial q_i} 
 \right. \nonumber \\
\left.  +  \frac {\partial^2 H}{\partial p_i \partial p_j } \frac {\partial H}
{\partial q_i } \frac {\partial H}{\partial q_j } \right) + 
o(\hbar^2)
\stackrel{\rm def}=\  H^3 + 3\hbar^2 \Delta H  + \hbar^2 Q +
o(\hbar^2).
  \ee
  To order $\hbar^2$, the products $H*H*H*H$ etc. are all 
expressed via $\Delta$ and $Q$,
  \be
  \label{Hnstar} 
\underbrace{H* \cdots *H}_n \ =
\nonumber \\ 
H^n + \hbar^2 \frac {n(n-1)}2 \Delta H^{n-2} 
+ \hbar^2 \frac {n(n-1)(n-2)}6 Q H^{n-3} 
+ o(\hbar^2)\ .
 \ee
Substituting (\ref{Hnstar}) into (\ref{WexpH}), we obtain
 \be
 \label{corrDelQ}
 \left[ e^{ - \beta \hat H} \right]_W  = \ e^{-\beta H} \left[ 1 + 
\frac {\beta^2 \hbar^2}2 \Delta - \frac {\beta^3 \hbar^2}6 Q +
o(\hbar^2) \right]\ .
 \ee
 Integrating by parts, one can derive
 \be
 \label{byparts}
 \int \beta Q e^{-\beta H} \prod_i \frac {dp_i dq_i}{2\pi}
 \ =\ 2\int \Delta e^{-\beta H} \prod_i \frac {dp_i dq_i}{2\pi}\ ,
  \ee
  which leads to the result (\ref{corrZHpq}).
  
  It is not difficult to generalize all this to the systems involving
fermion variables. We have
 \be
 \label{Ssled}
{\rm Tr} \{ (-1)^F e^{-\beta H} \}  = 
\int \prod_i \frac {dp_i dq_i}{2\pi} \prod_a d\bar\psi_a d\psi_a 
\left[ e^{ - \beta \hat H} \right]_W \ .
 \ee
 The star product of two functions on the phase space $(p_i,q_i; 
\bar\psi_a, \psi_a)$ is given, again, by the expression (\ref{starprod}), 
only we have to write the differential operator
  \be
  \label{fulloper}
  \frac {\partial^2}{\partial\Psi_a \partial \bar \psi_a}
-  \frac {\partial^2}{\partial\psi_a \partial \bar \Psi_a}
+ i\left(  \frac {\partial^2}{\partial q_i \partial P_i}
-  \frac {\partial^2}{\partial Q_i \partial p_i} \right) 
 \ee
 in the exponent. 
 Repeating all the steps of the derivation above, we are led to the result 
(\ref{Ceccorr}), (\ref{Delta}).

\end{document}